\documentstyle[prl,aps,twocolumn,epsfig]{revtex}

\newcommand{\IP}{\mbox{I}\!\mbox{P}}
\newcommand{\Li}{{\rm Li}}
\newcommand{\GeV}{{\rm\,GeV}}

\begin{document}

\title{$O(\alpha_s)$ corrections to longitudinal and transverse $W$-bosons
in top quark decays}

\author{J.G.~K\"orner\thanks{Talk given at the ``Third Latin American
Symposium on High Energy Physics (SILAFAE-III)'', Cartagena, Columbia,
April 2--8, 2000. The presented work is done in collaboration with M.~Fischer,
S.~Groote and M.C.~Mauser}}
\address{Institut f\"ur Physik der Johannes-Gutenberg-Universit\"at,
Staudinger Weg 7, D--55099 Mainz, Germany}

\maketitle

\begin{abstract}
In this talk we consider the $O(\alpha_s)$ radiative corrections to
the decay of an unpolarized top quark into a bottom quark and a $W$ gauge
boson where the helicities of the $W$ are specified as transverse-plus,
transverse-minus and longitudinal. The $O(\alpha_s)$ radiative corrections
lower the normalized longitudinal rate $\Gamma_L/\Gamma$ by 0.8\%. We find
that the normalized transverse-plus rate $\Gamma_+/\Gamma$, which vanishes
at the Born term level for $m_b\rightarrow0$, receives radiative correction
contributions at the sub-percent level. Our results are discussed in the
light of recent measurements of the helicity content of the $W$ in top quark
decays by the CDF Collaboration.
\end{abstract}

\section{Introduction} 
The CDF Collaboration has recently published the results of a first
measurement of the helicity content of the $W$ gauge boson in top quark
decays~\cite{cdf}. Their results are
\begin{eqnarray}
\Gamma_L/\Gamma&=&0.91\pm0.37(stat)\pm0.13(syst)\qquad\label{CDFlong}\\
\Gamma_+/\Gamma&=&0.11\pm0.15\label{CDFtrans+}
\end{eqnarray}
where $\Gamma_L$ and $\Gamma_+$ denote the rates into the longitudinal and
transverse-plus polarization state of the $W$-boson and $\Gamma$ is the total
rate. 

The errors on this measurement are still ra\-ther large but will be much
reduced when larger data samples become available in the future from TEVATRON
RUN II, and, at a later stage, from the LHC. Optimistically the measurement
errors can eventually be reduced to the $(1-2)\%$ level~\cite{willenbrock}.
If such a level of accuracy can in fact be reached it is important to discuss
the radiative corrections to the different helicity rates~\cite{FGKLM,FGKM}
considering the fact that the $O(\alpha_s)$ radiative corrections to the
total width $\Gamma$ are rather large
($\approx-9.4\%$)~\cite{c14,c15,c16,c17,c18}. The transverse-plus rate
$\Gamma_+$ is particularly interesting in this regard. Simple helicity
considerations show that $\Gamma_+$ vanishes at the Born term level in the
$m_b=0$ limit. A nonvanishing transverse-plus rate could arise from i)
$m_b\ne0$ effects, ii) $O(\alpha_s)$ radiative corrections due to gluon
emission, or from iii) non Standard Model $t\rightarrow b$ currents. As we
shall show in this talk the $O(\alpha_s)$ and the $m_b\ne 0$ corrections to
the transverse-plus rate occur only at the sub-percent level. It is safe to
say that if top quark decays reveal a violation of the Standard Model $(V-A)$
current structure which exceeds the $1\%$ level, the violations must have a
non Standard Model origin.

\section{Angular decay distribution}
Let us begin by writing down the angular decay distribution for the decay
process $t\rightarrow X_b+W^+$ followed by $W^+\rightarrow l^++\nu_l$ (or by
$W^+\rightarrow\bar q+q$). For unpolarized top decay the angular decay
distribution is determined by two transverse components (transverse-plus and
transverse-minus) and the longitudinal component of the $W$-boson. One
has~\cite{FGKLM}
\begin{eqnarray}\label{ang}
\frac{d\Gamma}{d\cos\theta}&=&\frac38(1+\cos\theta)^2\Gamma_+
  +\frac38(1-\cos\theta)^2\Gamma_-\nonumber\\&&\qquad
  +\frac34\sin^2\theta\ \Gamma_L.
\end{eqnarray}
Integrating over $\cos\theta$ one recovers the total rate
\begin{equation}
\Gamma=\Gamma_++\Gamma_-+\Gamma_L.
\end{equation}

\begin{figure}
\centering\leavevmode\psfig{file=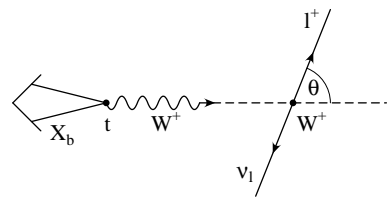,width=5cm}
  \caption{\label{fig1}Definition of the polar angle $\theta$}
\end{figure}

We describe the angular decay distribution in cascade fashion, i.e.\ the
polar angle $\theta$ is measured in the $W$ rest frame where the lepton pair
or the quark pair emerges back-to-back. The angle $\theta$ denotes the polar
angle between the $W^+$ momentum direction and the antilepton $l^+$ (or the
antiquark $\bar q$) (see Fig.~1). The various contributions in (\ref{ang})
are reflected in the shape of the lepton energy spectrum in the rest frame of
the top quark. From the angular factors in (\ref{ang}) it is clear that the
contribution of $\Gamma_+$ makes the lepton spectrum harder while $\Gamma_-$
softens the spectrum where the hardness or softness is gauged relative to the
longitudinal contribution. The only surviving contribution in the forward
direction $\theta=0$ comes from $\Gamma_+$. The fact that $\Gamma_+$ is
predicted to be quite small implies that the lepton spectrum will be soft.
The CDF measurement of the helicity content of the $W^+$ in top decays was in
fact done by fitting the values of the helicity rates to the shape of the
lepton's energy spectrum.

The angular decay distribution of the antitop decay
$\bar t\rightarrow X_{\bar b}+W^-$ followed by $W^-\rightarrow l^-+\bar{\nu_l}$
(or by $W^-\rightarrow q+\bar q$) can be obtained from the angular decay
distribution (\ref{ang}) by the substitution
$(1+\cos\theta)^2\leftrightarrow(1-\cos\theta)^2$. The polar angle $\theta$
is now defined with regard to the lepton $l^-$ (or the quark) direction.

\begin{figure}
\centering\leavevmode
\put(5,5){a)}\raise22pt\hbox{\psfig{file=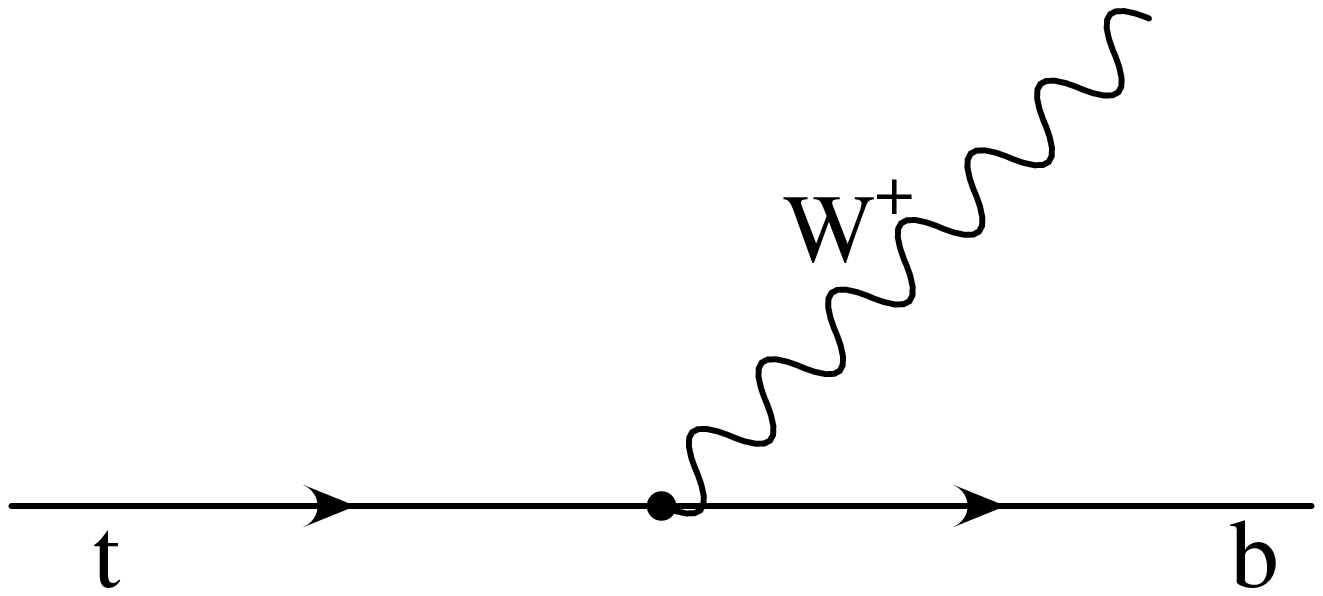,width=3cm}}\kern12pt
\put(5,5){b)}\raise4.5pt\hbox{\psfig{file=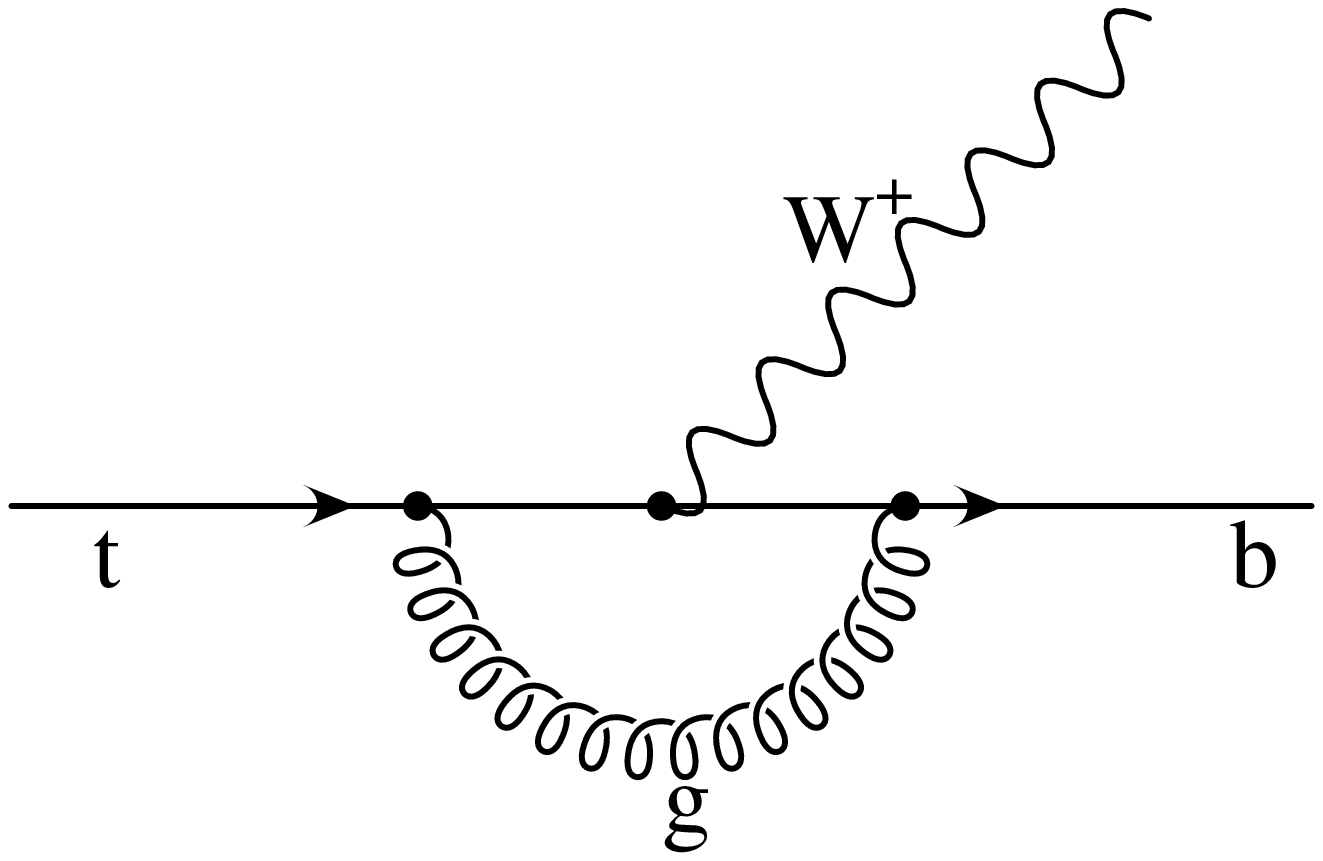,width=3cm}}\newline
\put(5,5){c)}\hbox{\psfig{file=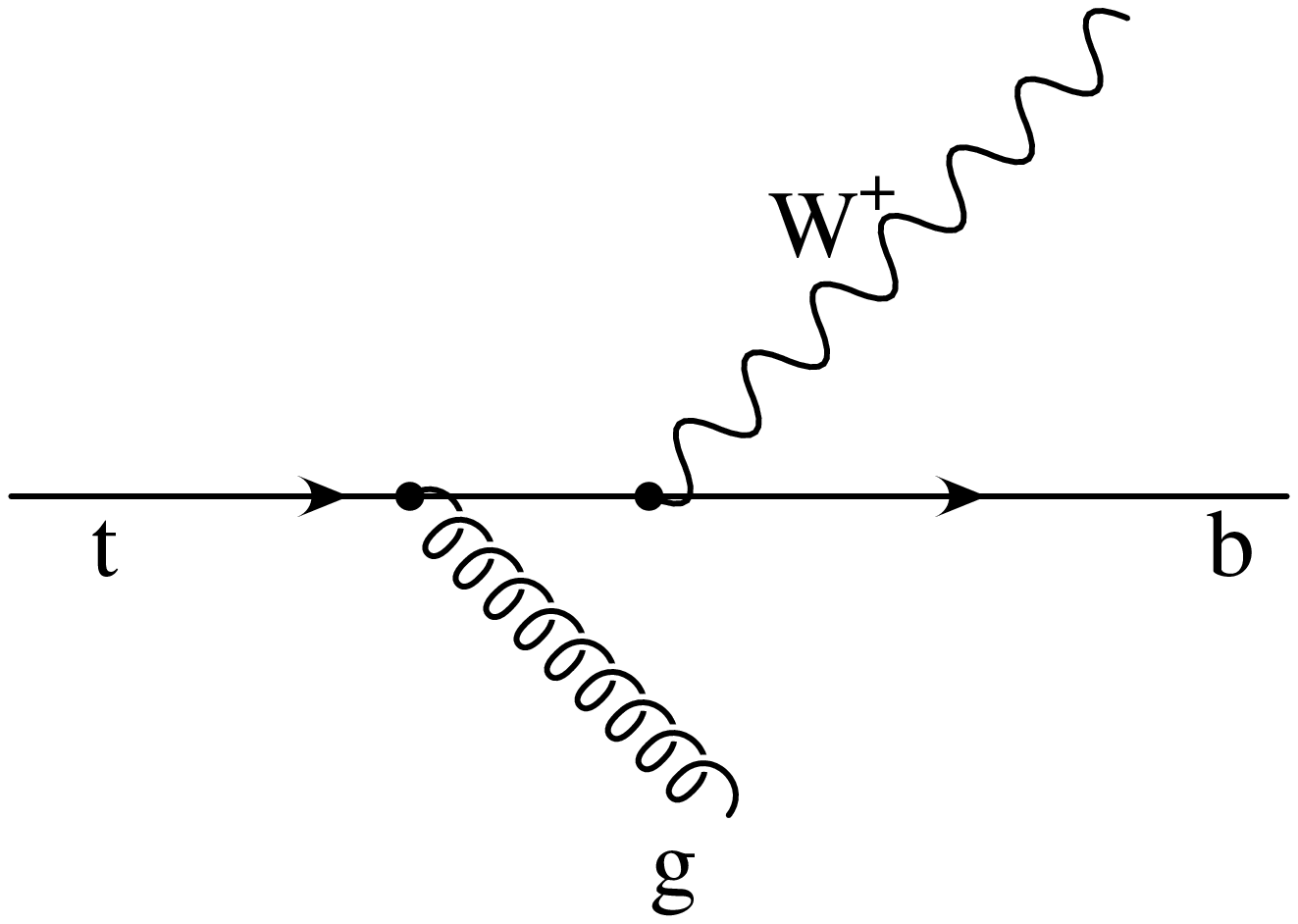,width=3cm}}\kern12pt
\put(5,5){d)}\hbox{\psfig{file=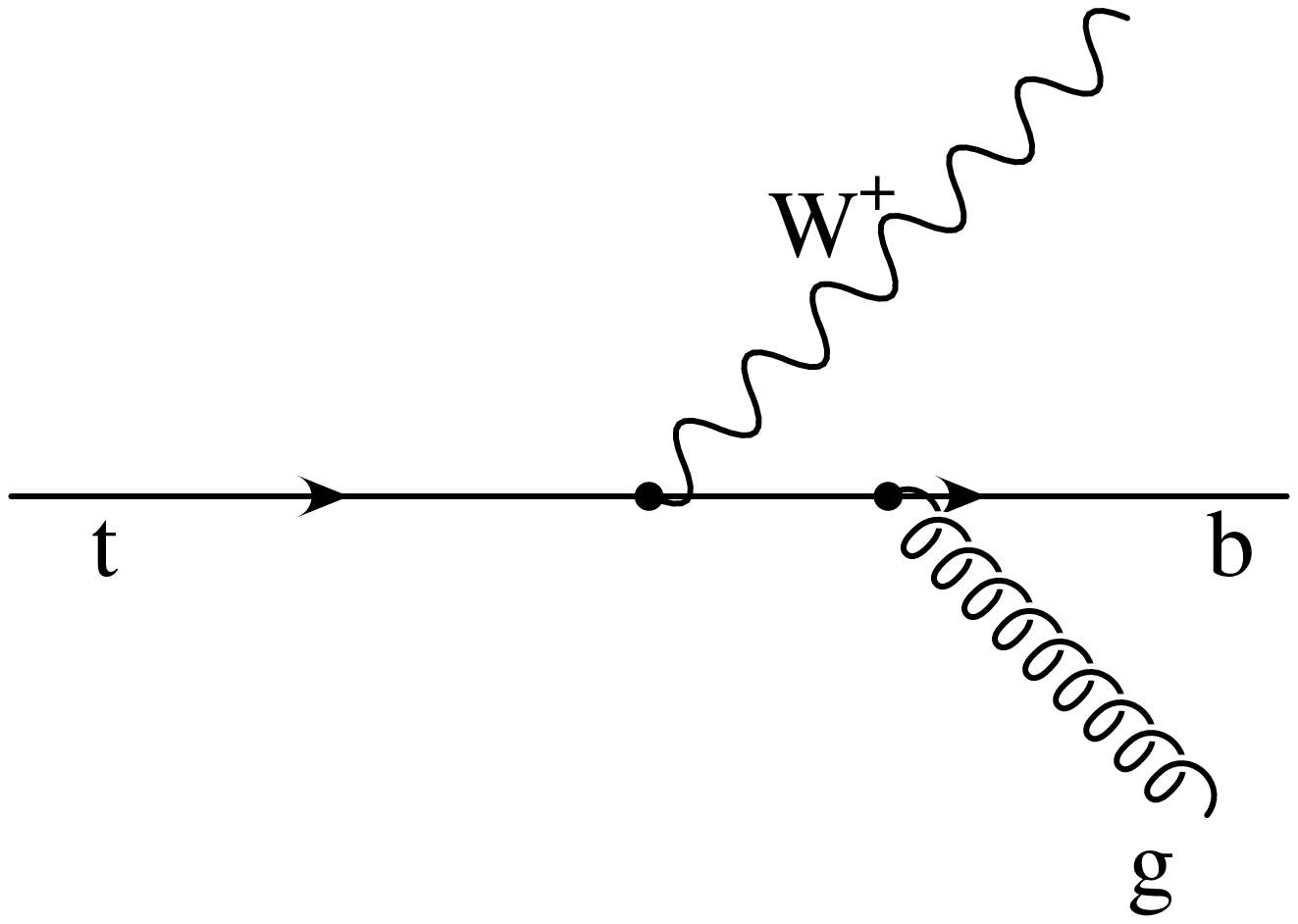,width=3cm}}
\caption{\label{fig2}Leading order Born term contribution (a) and $O(\alpha_s)$
  contributions (b,c,d) to $t\rightarrow b+W^+$.}
\end{figure}

\section{Born term results}
The Born term contribution is shown in Fig.~\ref{fig2}a. We use the scaled
masses $x=m_W/m_t$, $y=m_b/m_t$ and the K\"all\'en-type function
$\lambda=1+x^4+y^4-2x^2y^2-2x^2-2y^2$. Including the full $m_b$-dependence,
the Born term rate is given by
\begin{eqnarray}\label{Born}
\Gamma_0&=&\frac{G_Fm_W^2m_t}{8\sqrt2\pi}|V_{tb}|^2\sqrt\lambda
  \times\nonumber\\&&\frac{(1-y^2)^2+x^2(1-2x^2+y^2)}{x^2}.
\end{eqnarray}
The partial helicity rates are given in terms of the Born term rate. One has
\begin{eqnarray}
\Gamma_L/\Gamma_0&=&\frac{(1-y^2)^2-x^2(1+y^2)}
  {(1-y^2)^2+x^2(1-2x^2+y^2)}=\frac1{1+2x^2}+\ldots\nonumber\\
\Gamma_+/\Gamma_0&=&\frac{x^2(1-x^2+y^2-\sqrt\lambda)}
  {(1-y^2)^2+x^2(1-2x^2+y^2)}\nonumber\\
  &=&y^2\frac{2x^2}{(1-x^2)^2(1+2x^2)}+\ldots\label{Born+}\\
\Gamma_-/\Gamma_0&=&\frac{x^2(1-x^2+y^2+\sqrt\lambda)}
  {(1-y^2)^2+x^2(1-2x^2+y^2)}=\frac{2x^2}{1+2x^2}+\ldots\nonumber
\end{eqnarray}
In Eqs.~(\ref{Born+}) we have also listed the leading components of the small
$y^2$ expansion of the Born term rate ratios. As the second of
Eqs.~(\ref{Born+}) shows and as already remarked on before, the
transverse-plus Born term contribution vanishes in the $m_b=0$ limit. Note
that the leading contribution to the transverse-plus rate is proportional to
$(m_b/m_t)^2$ and not proportional to $(m_b/m_W)^2$ as stated in
Ref.~\cite{cdf}.

The $m_b\ne0$ effects are quite small. Using $m_t=175\GeV$, $m_W=80.419\GeV$,
and a pole mass of $m_b=4.8\GeV$~\cite{pivovarov}, one finds that $\Gamma_0$
and $\Gamma_L/\Gamma_0$ decrease by $0.266\%$ and $0.091\%$, resp.\ when going
from $m_b=0$ to $m_b=4.8\GeV$, and $\Gamma_-/\Gamma_0$ increases by $0.094 \%$.
The leakage into the transverse-plus rate ratio $\Gamma_+/\Gamma_0$ from
bottom mass effects is a mere $0.036\%$.  

\section{$O(\alpha_s)$ radiative corrections}
The $O(\alpha_s)$ corrections are determined by the one-loop vertex correction
shown in Fig.~\ref{fig2}b and the gluon emission graphs shown in
Figs.~\ref{fig2}c and \ref{fig2}d. The one-loop results have already been
known for quite some time~\cite{schilcher,gounaris} and will not be discussed
any further in this talk.

We do want to make a few technical remarks about how the tree-graph integration
was done. We use a gluon mass to regularize the IR singularity. Concerning the
collinear singularity we have kept the full bottom mass dependence in our
calculation and have only set the bottom mass to zero at the very end. We have
thus effectively used a mass regulator to regularize the collinear singularity.
The tree-graph integration has to be done over two-dimensional phase space. As
phase space variables we use the gluon energy $k_0$ and the $W$ energy $q_0$.
The IR behaviour of the hadronic tree-graph matrix element
$W_{\mu\nu}(q_0,k_0)$ was improved by subtracting from it the soft-gluon
contribution $G_{\mu\nu}(q_0,k_0)$ which was then added again according to the
prescription
\begin{eqnarray}
W_{\mu\nu}(q_0,k_0)&=&(W_{\mu\nu}(q_0,k_0)-G_{\mu\nu}(q_0,k_0))\nonumber\\&&
  \qquad+G_{\mu\nu}(q_0,k_0)
\end{eqnarray}
The first piece $(W_{\mu\nu}(q_0,k_0)-G_{\mu\nu}(q_0,k_0))$ has thereby been
rendered IR finite and can be integrated without a gluon mass regulator which
considerably simplifies the phase space integration. The IR singularity resides
in the soft gluon piece $G_{\mu\nu}(q_0,k_0)$ which is, however, simple and
universal and can be easily integrated. In fact the soft gluon contribution
factorizes into the Born term contribution $B_{\mu\nu}$ and a universal soft
gluon factor $S(q_0,k_0)$ according to
\begin{equation}\label{sga}
G_{\mu\nu}(q_0,k_0)=B_{\mu\nu}\cdot S(q_0,k_0)\cdot\alpha_s.
\end{equation}
The Born term contribution $B_{\mu\nu}$ is given by
\begin{equation}\label{bornspur}
B^{\mu\nu}=8(p_t^\nu p_b^\mu+p_t^\mu p_b^\nu-g^{\mu\nu}p_t\cdot p_b
  +i\epsilon^{\mu\nu\alpha\beta}p_{b,\alpha}p_{t,\beta}),
\end{equation}
while the soft-gluon factor in (\ref{sga}) has the standard form
\begin{equation}
S(q_0,k_0)=\frac{m_t^2}{(p_tk)^2}-\frac{2p_tp_b}{(p_tk)(p_bk)}
  +\frac{m_b^2}{(p_bk)^2}.
\end{equation}
Note that the tensor structure carrying the spin information of the produced
$W$-boson has been factored out and is now entirely contained in the Born term
factor $B_{\mu\nu}$. Since the Born term factor does not depend on the phase
space variables, the phase space integration needs to be done only with respect
to the soft gluon factor $S(q_0,k_0)$ and thus needs to be done only once
irrespective of the polarisation of the $W$-boson. Needless to say that this is
a very welcome simplifying feature of the above subtraction procedure.

This is different for the phase space integration of the IR-finite piece
$(W_{\mu\nu}(q_0,k_0)-G_{\mu\nu}(q_0,k_0))$ where the integration has to
be done separately for each polarization state of the $ W $-boson. To do the 
necessary two-dimensional phase space integrations in analytical form is
somewhat involved. In particular the integrations are more difficult than
those needed for the total rate calculation which has already been done some
time ago~\cite{c14,c15,c16,c17,c18}. The complicating feature can be best
appreciated by discussing how one obtains the various helicity components of
the $W$-boson.
  
We chose to use covariant projectors to pro\-ject out the various helicity
components of the $W$-boson. For the total rate one has the familiar form
\begin{equation}\label{projtot} 
\IP^{\mu \nu}_{U+L}=-g^{\mu\nu}+\frac{q^{\mu}q^{\nu}}{m_W^2}
\end{equation}
where we have added the label $(U+L)$ for added emphasis. The projector
(\ref{projtot}) can be seen to reduce to the appropiate three-dimensional form
$\delta_{ij}$ in the $W$ rest system. The longitudinal helicity rate is
obtained with the help of the projector
\begin{equation}
\IP^{\mu \nu}_L=\frac{m_W^2}{m_t^2}\frac1{|\vec q\,|^2}
  \Big(p_t^\mu-\frac{p_t\cdot q}{m_W^2}q^\mu\Big)
  \Big(p_t^\nu-\frac{p_t\cdot q}{m_W^2}q^\nu\Big).
\end{equation}
The transverse-plus and transverse-minus helicity projectors are obtained in
an indirect way by first considering the sum and the difference of the
transverse-plus and transverse-minus projectors. The sum is labelled by the
index $U$ (unpolarized-transverse), and the corresponding projector is obtained
from $\IP^{\mu\nu}_U=\IP^{\mu\nu}_{U+L}-\IP^{\mu\nu}_L$. The difference of the
transverse-plus and trans\-verse-minus helicity projectors is labelled by the
index $F$ (forward-backward) projector which is given by
\begin{equation}
\IP^{\mu\nu}_F=-\frac1{m_t}\frac1{|\vec q\,|}
  i\epsilon^{\mu\nu\alpha\beta}p_{t,\alpha}q_\beta.
\end{equation}
The transverse-plus and transverse-minus projectors are thus determined by the
linear combinations
$\IP^{\mu\nu}_\pm=\frac12(\IP^{\mu\nu}_U\pm\IP^{\mu\nu}_F)$.

The inverse powers of the magnitude $|\vec q\,|=\sqrt{q_0^2-m_W^2}$ of the
three-momentum of the $W$-boson appear in the projectors for normalization
reasons. It is mainly because of the additional factors of $|\vec q\,|^{-n}$ in
the helicity rates that makes their phase space integration technically more
involved than the total rate integration since new classes of phase space
integrals appear.       

Our final results are presented for the $m_b=0$ limit. We add together the
Born term, the one-loop and the tree-graph contribution. The results are taken
from Ref.~\cite{FGKLM}. We divide out the total $m_b=0 $ Born term rate and
denote the scaled rates $\hat\Gamma_i=\Gamma_i/\Gamma_0$ ($i=U+L,L,+,-$) by a
hat symbol. The two transverse helicity rates $i=+,-$ are obtained from the
$i=U,F$ rates given in~\cite{FGKLM} by taking the linear combinations
$\Gamma_\pm=\frac12(\Gamma_U\pm\Gamma_F)$ as discussed before. One has
\begin{eqnarray}
\lefteqn{\hat\Gamma\ =\ 1+\frac{\alpha_s}{2\pi}C_F
  \frac{x^2}{(1-x^2)^2(1+2x^2)}\times}\nonumber\\&&\kern-7pt
  \Bigg\{\frac{(1-x^2)(5+9x^2-6x^4)}{2x^2}
  -4(1+x^2)(1-2x^2)\ln(x)\nonumber\\[-3pt]&&
  -\frac{(1-x^2)^2(5+4x^2)}{x^2}\ln(1-x^2)\nonumber\\&&\kern-3pt
  -\frac{4(1-x^2)^2(1+2x^2)}{x^2}
  \left(\ln(x)\ln(1-x^2)+\frac{\pi^2}6\right)\nonumber\\[-3pt]&&\kern-7pt
  -\frac{8(1-x^2)^2(1+2x^2)}{x^2}\left(\Li_2(x)+\Li_2(-x)\right)\Bigg\}\\
\lefteqn{\hat\Gamma_L\ =\ \frac1{1+2x^2}+\frac{\alpha_s}{2\pi}C_F
  \frac{x^2}{(1-x^2)^2(1+2x^2)}\times}\nonumber\\&&\kern-7pt
  \Bigg\{\frac{(1-x^2)(5+47x^2-4x^4)}{2x^2}
  -\frac{(1+5x^2+2x^4)}{x^2}\ \frac{2\pi^2}3\nonumber\\[-3pt]&&
  +16(1+2x^2)\ln(x)-\frac{3(1-x^2)^2}{x^2}\ln(1-x^2)\nonumber\\&&
  -2(1-x)^2\frac{2-x+6x^2+x^3}{x^2}\ln(1-x)\ln(x)\nonumber\\&&
  -2(1+x)^2\frac{(2+x+6x^2-x^3)}{x^2}\ln(x)\ln(1+x)\nonumber\\&&
  -2(1-x)^2\frac{(4+3x+8x^2+x^3)}{x^2}\Li_2(x)\nonumber\\[-3pt]&&\kern-7pt
  -2(1+x)^2\frac{(4-3x+8x^2-x^3)}{x^2}\Li_2(-x)\Bigg\}\\
\lefteqn{\hat{\Gamma}_+\ =\ \frac{\alpha_s}{2\pi}C_F
  \frac{x^2}{(1-x^2)^2(1+2x^2)}\times}\nonumber\\&&\kern-7pt
  \Bigg\{-\frac12(1-x)(25+5x+9x^2+x^3)\nonumber\\[-3pt]&&
  +(7+6x^2-2x^4)\frac{\pi^2}3-2(5+7x^2-2x^4)\ln(x)\nonumber\\[3pt]&&
  -2(1-x^2)(5-2x^2)\ln (1+x)\nonumber\\[3pt]&&
  -\frac{(1-x)^2}{x}(5+7x^2+4x^3)\ln(x)\ln(1-x)\nonumber\\&&
  +\frac{(1+x)^2}{x}(5+7x^2-4x^3)\ln(x)\ln(1+x)\nonumber\\&&
  -\frac{(1-x)^2}{x}(5+7x^2+4x^3)\Li_2(x)\\[-3pt]&&\kern-7pt
  +\frac1x(5\!+\!10x\!+\!12x^2\!+\!30x^3\!-\!x^4\!-\!12x^5)\Li_2(-x)\Bigg\}
  \nonumber\\
\lefteqn{\hat{\Gamma}_-\ = \ \frac{2x^2}{1+2x^2}+\frac{\alpha_s}{2\pi}C_F
  \frac{x^2}{(1-x^2)^2(1+2x^2)}\times}\nonumber\\&&\kern-7pt
  \Bigg\{-\frac12(1-x)(13+33x-7x^2+x^3)\nonumber\\[-3pt]&&\kern-3pt
  +(3+4x^2-2x^4)\frac{\pi^2}3-2(5+7x^2-2x^4)\ln(x)\nonumber\\&&
  -2\frac{(1-x^2)^2(1+2x^2)}{x^2}\ln(1-x)\nonumber\\&&
  -2\frac{(1-x^2)(1-4x^2)}{x^2}\ln(1+x)\nonumber\\&&
  -\frac{(1-x)^2}{x}(5+7x^2+4x^3)\ln(x)\ln(1-x)\nonumber\\&&
  +\frac{(1+x)^2}{x}(5+7x^2-4x^3)\ln(x)\ln(1+x)\nonumber\\&&
  -\frac{(1-x)^2}{x}(5+3x)(1+x+4x^2)\Li_2(x)\\[-3pt]&&\kern-7pt
  +\frac1x(5+2x+12x^2+6x^3-x^4-4x^5)\Li_2(-x)\Bigg\}\nonumber
\end{eqnarray}
The expressions for the rates contain the usual logarithmic and dilogarithmic
factors that appear in one-loop radiative correction calculations. As expected,
the transverse-plus rate becomes non-zero only at the $O(\alpha_s)$ level. Note
that the entire $O(\alpha_s)$ contribution to the transverse-plus rate comes
from the tree-graph contribution since the one-loop graph does not contribute
to the transverse-plus rate for chirality reasons. As expected, the
longitudinal rate becomes dominant in the high energy limit as
$m_t\rightarrow\infty$ for both the Born term and the $O(\alpha_s)$
corrections.

\section{Numerical results}
We are now in a position to discuss our numerical results. Our input values
are $m_t=175\GeV $ and $m_W=80.419\GeV$, as before. For the strong coupling
constant we use $\alpha_s(m_t)=0.107$ which was evolved downward from
$\alpha_s(m_Z)=0.1175$. Our numerical results are presented in terms of the
hatted helicity rates $\hat\Gamma_i=\Gamma_i/\Gamma_0$ ($i=U+L,L,+,-$)
introduced in Sec.~4. In order to be able to quickly assess the percentage
changes induced by the $ O(\alpha_s) $ corrections, we have factored out the
Born term helicity rates (when applicable) from the $O(\alpha_s)$ results.
One has
\begin{eqnarray}
 \hat{\Gamma}_{\phantom{+}} & = & \phantom{0.703 (} 1-0.0938, \\
 \hat{\Gamma}_{L}           & = & 0.703 (1-0.104), \\
 \hat{\Gamma}_{+}           & = & \phantom{0.703 (} 0.001018, \\
 \hat{\Gamma}_{-}           & = & 0.297 (1-0.0720),
\end{eqnarray}

The radiative corrections to the longitudinal and transverse-minus rates are
sizeable where the radiative correction to the longitudinal rate is larger.
The radiative corrections lower the normalized longitudinal rate
$\Gamma_L/\Gamma$ by $0.8\%$ and increase the normalized transverse-minus
rate $\Gamma_-/\Gamma$ by $2.0\%$. The radiative correction to the
trans\-verse-plus rate is quite small. For the normalized transverse-plus rate
$\Gamma_+/\Gamma$ we obtain a mere $0.10\%$ which is only marginally larger
than the value of $0.036\%$ obtained from the Born term level $m_b\ne0$
effects discussed in Sec.~3.

\section{Summary and conclusions}
In this talk we have presented results on the $O(\alpha_s)$ radiative
corrections to the three helicity rates in unpolarized top quark decays which
can be determined from doing an angular analysis or from an analysis of the
shape of the lepton spectrum. While the radiative corrections to the
unnormalized transverse-minus and longitudinal rate are sizable
($\approx 6-10\%$), the radiative corrections to the normalized helicity rates
are smaller ($\approx 1-2\%$). The radiative correction to the transverse-plus
rate is very small. The measurements of the helicity rates by the CDF
Collaboration can be seen to be fully compatible with the predictions of the
Standard Model. The errors on these measurements are, however, too large to
allow one to meaningfully compare the present measurements with quantum
effects brought in by QCD radiative corrections. There is hope that this will
change in the future.

\end{document}